\documentclass[intlimits,twoside,a4paper]{article}
\usepackage{graphicx}
\usepackage{wrapfig}
\usepackage{xcolor}
\usepackage[T2A]{fontenc}
\usepackage[cp1251]{inputenc}
\usepackage[eqsecnum]{cmpj3}

\usepackage{bm}

\issue{2025}{28}{1}{13802}
\doinumber{10.5488/CMP.28.13802}

\title[Thermodynamics of asymmetric valency aqueous salts]
{Effect of variable relative permittivity on the thermodynamics of asymmetric valency
aqueous salts}

\author[A. O. Qui\~{n}ones, Z. Abbas, C. W. Outhwaite,  L. B. Bhuiyan]{
A. O. Qui\~{n}ones\orcid{0000-0003-2982-8971}\refaddr{label1}, Z. Abbas\orcid{0000-0003-1741-1925}\refaddr{label2}, C. W. Outhwaite\orcid{0000-0002-8825-2127}\refaddr{label3},
L. B. Bhuiyan\orcid{0000-0002-4574-173X}\refaddr{label1} }

\addresses{
\addr{label1}Laboratory of Theoretical Physics, Department of
Physics, University of Puerto Rico, 17 Avenida Universidad, STE 1701, San Juan, Puerto
Rico 00925-2537, USA
\addr{label2}Department of Chemistry and Molecular Biology, University of Gothenburg,
Kening{\aa}rden 4, SE-41296, Gothenburg, Sweden
\addr{label3}School of Mathematical and Physical Sciences, University of Sheffield,
Sheffield S3 7RH, UK
}

\date{Received January 20, 2025, in final form March 6, 2025}

\begin{document}

\maketitle
\begin{abstract}
Experimentally determined empirical formulae for the
concentration dependent relative permittivity of aqueous solutions
of MgCl$_{2}$ and NiCl$_{2}$ are utilized to calculate the osmotic
coefficient and the mean activity coefficient of these salts for a
range of concentrations. The systems are modelled using the
primitive model of electrolytes and analyzed using
the symmetric Poisson-Boltzmann theory, the modified Poisson-Boltzmann
theory, the mean spherical approximation, and the Monte Carlo
simulations. Generally, the mean spherical approximation and the
modified Poisson-Boltzmann theory reproduce the \emph{benchmark}
simulation data well up to~$\sim $1.6~mol/dm$^{3}$ or more
in many instances, while the symmetric Poisson-Boltzmann results
show discrepancies starting from~$\sim $0.25~mol/dm$^{3}$. Both the
simulations and the theories tend to deviate from the corresponding
experimental results beyond $\sim $1 mol/kg.

\printkeywords
\end{abstract}

\section{Introduction}

Osmotic and activity coefficients are two of the more
important quantities in the thermodynamic description of charged
fluids. From a practical point of view such data are relevant for many
biological systems and for chemical industrial processes involving fluids
\cite{Harned,Levin,Doug,Chersty}. Over the last  seven decades or so
a lot of experimental effort has been expended in chronicling such
measurements (see for example,~\cite{Robinson,Barthel} and
references therein).

    On the theoretical and numerical simulation front, the availability of
powerful computational technology has been a boon to research
in this field.
The development of formal statistical mechanical theories of electrolytes in
general has been greatly aided by numerical computer simulations of physical
models, such data often being likened to as the \emph{benchmark}.
We quote here some of the more pertinent literature in this regard \cite{Outh,Hansen,
Vlachy,McQuarrie,Friedman,Blum,Bhuiyan}.

    A widely used model of electrolytes and molten salts is a primitive model
(PM). The ions are modelled as rigid spheres of arbitrary radii with an embedded
charge of arbitrary valency at the centre of each ion. When the ionic sizes are
the same, it is a restricted primitive model (RPM). It is noteworthy that the
underlying model of the classical Debye-H\"{u}ckel (DH) theory \cite{Debye} is the
RPM with  vanishing radii. The conceptual simplicity and the ease of implementation
of the DH makes it a popular first approximation in analyzing experimental data. The
DH is a mean field theory and simulations supported by formal theories have shown
over the years that its two basic deficiencies are the neglect of (i) fluctuation
potential (inter-ionic correlations), and (ii) ionic exclusion volume effects. This
makes the theory relevant only for monovalent systems at rather low concentrations.
The PM (and the RPM), on the other hand, have proved useful in describing structure
and thermodynamics of electrolytes at solution concentrations~\cite{Outh,Vlachy,Abbas,
Abbas2,Quinones,Outh2,Outh2ol,Bhuiyan2}.

    Traditionally, in the simulations and in the formal treatments of electrolytes,
the relative permittivity~$\varepsilon _{r}$ (dielectric constant) is held fixed.
Experiments, however, have revealed that the $\varepsilon _{r}$ is not constant
but is influenced by both the concentration and the temperature
\cite{Robinson,Barthel}. Some limited theoretical works on variable $\varepsilon _{r}$
were attempted using the DH theory \cite{Huckel,Shilov} and the MSA \cite{Triolo, Simonin,
Fawcett, Tikanen}. More recently, Abbas and Ahlberg \cite{Abbas3} have done MC simulations
for hydrogen halide and some alkali halides to calculate the osmotic coefficient $\phi $
and the mean ionic activity coefficients $y_{\pm}$ using concentration dependent and
temperature dependent $\varepsilon _{r}$. In a previous paper \cite{Quinones2}, 
we applied the symmetric Poisson-Boltzmann (SPB), the modified
Poisson-Boltzmann (MPB), and the MSA theories at the parameters of Abbas and Ahlberg's
\cite{Abbas3} simulations with encouraging results. The MSA, the MPB, and to a lesser extent the SPB
could reproduce the MC data for $\phi $ and $y_{\pm}$ to a high degree of accuracy.

    At a fundamental level, the simulations and the theories are based on the McMillan-Mayer (MM)
formalism. The solvent is likened to a structureless continuum so that the ionic pair-potential
is essentially an effective inter-ionic potential that incorporates solvent effects. The solution
permittivity at infinite solute dilution is the permittivity of the pure solvent. It thus stands
to reason, as Friedman \cite{Friedman2} showed, that the effective potential should be concentration
and temperature dependent. Friedman's analysis further demonstrated how non-pairwise terms
in the effective potential lead to a concentration dependent $\varepsilon _{r}$. For a more
detailed discussion of this point we refer the reader to \cite{Quinones2}. Although analytical efforts to
obtain a closed expression for $\varepsilon _{r}$ have proved difficult \cite{Levy,Valisko,
Gavish,Adar}, progress can be made by making recourse to experimental data. Indeed, this was
done in the MC simulations of Abbas and Ahlberg \cite{Abbas3} and in~\cite{Quinones2}. For each electrolyte
treated, an empirical formula was used for $\varepsilon _{r}$ based on the measured value of
$\varepsilon _{r}$ over a range of concentrations \cite{Barthel}. Such a scheme
is also fairly straightforward to implement in both simulations and in the SPB, MPB, and
MSA theories.

    The electrolytes studied in \cite{Quinones2} were all 1:1 symmetric valency systems. In the present
work we  extend the procedure to higher valency asymmetric 2:1 cases --- MgCl$_{2}$
and NiCl$_{2}$. The empirical formulae for~$\varepsilon _{r}$ as function of concentration
$c$, that we  use for these salts, are as follows:
\begin{equation}
\varepsilon _{r}=78.36 - 34c + 10.9c^{2} - 1.5c^{3}
\label{eq1.1}
\end{equation}
MgCl$_{2}$ \cite{Barthel}, and
\begin{equation}
\varepsilon _{r}=78.36 - 25c + 8.7c^{2}
\label{eq1.2}
\end{equation}
for NiCl$_{2}$ \cite{Hasted}.
\begin{figure}[!t]
		\begin{center}
	\centerline{\includegraphics[width=0.4\textwidth]{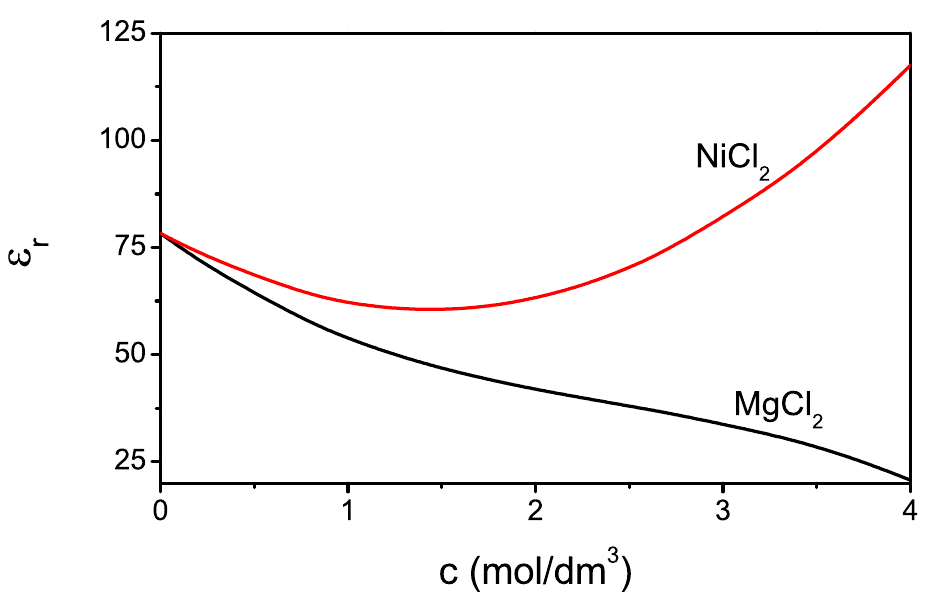}}
	\caption{(Colour online) Variation of the relative permittivity $\varepsilon _{r}$
		as a function of concentration $c$~{[in~(mol/dm$^{3}$)]} for MgCl$_{2}$ and NiCl$_{2}$}
		\label{fig1}
\end{center}
\end{figure}

 There is another important  difference between \cite{Quinones2} and the present work.
For instance, in \cite{Quinones2}, the $\varepsilon _{r}$ for each of the electrolytes decreased with
increasing concentration --- the phenomenon known as \emph{dielectric decrement} in the
literature. By contrast, as can be seen in figure~\ref{fig1}, although the $\varepsilon _{r}$ decreases
with $c$ for MgCl$_{2}$, the $\varepsilon _{r}$ for NiCl$_{2}$ initially decreases, reaches
a minimum before increasing at higher concentrations. Apart from comparing the SPB, MPB,
and MSA results with that of the MC simulations, we  also compare both simulations
and the theories with available experimental $\phi $ and $y_{\pm }$ data for MgCl$_{2}$
and NiCl$_{2}$ \cite{Lobo,Stokes,Stokes2,Stokes3}.

\section{Model and Methods}

\subsection{Model}

The PM used here is a two component system appropriate for a
single electrolyte. The ions are mimicked by charged hard spheres of relevant
sizes and charges in a dielectric medium of concentration dependent $\varepsilon _{r}$.
The interionic pair potential is given by
\begin{equation}
u_{ij}(r) = \left\{ \begin{array}{cc}
\infty
~~~~~~~~~~~~~~~~~~~~~~~~~~~~~~~~~ r  < a_{ij}  \\
e^{2}Z_{i}Z_{j}/(4\piup \varepsilon _{0}\varepsilon _{r} r)~~~~~~ r > a_{ij} ,
\end{array}\right.
\end{equation}
where $a_{ij}=a_{i}+a_{j}$, $Z_{s}$, $a_{s}$ are the valency and radius of ionic species
$s$ respectively, $e$ is the proton charge, and $\epsilon _{0}$ is the vacuum permittivity.

\subsection{Methods}
\begin{figure}[!t]
	\begin{center}
		\centerline{\includegraphics[width=0.35\textwidth]{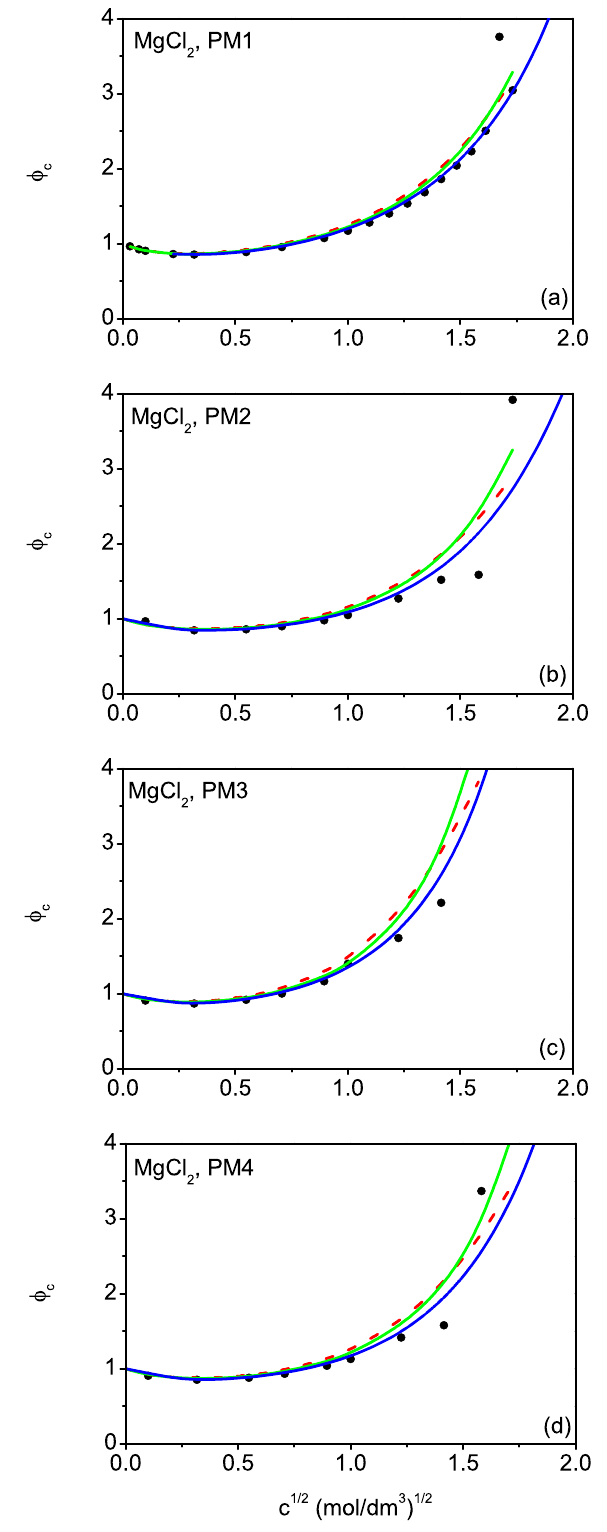}}
		\caption{(Colour online) MPB, SPB, MSA, and MC osmotic coefficient $\phi _{c}$
			(in molar scale) as function of square root of concentration $c^{1/2}$ (in (mol/dm$^{3}$)$^{1/2}$)
			for MgCl$_{2}$ using the PM with a concentration dependent dielectric constant $\varepsilon _{r}$.
			In all of the sub-figures, dashed red line SPB, solid green line MPB, solid blue line MSA, black filled circles MC;
			{\bf (a)} PM1, $a_{\text{Mg}^{2+}}$ = 2.95 $\cdot$ 10$^{-10}$~m, $a_{\text{Cl}^{-}}$ = 1.81 $\cdot$ 10$^{-10}$~m, fixed $\varepsilon _{r}$,
			{\bf (b)} PM2, $a_{\text{Mg}^{2+}}$ = 2.95 $\cdot$ 10$^{-10}$~m, $a_{\text{Cl}^{-}}$ = 1.81 $\cdot$ 10$^{-10}$~m, variable $\varepsilon _{r}$,
			{\bf (c)} PM3, $a_{\text{Mg}^{2+}}$ = 3.50 $\cdot$ 10$^{-10}$~m, $a_{\text{Cl}^{-}}$ = 1.81 $\cdot$ 10$^{-10}$~m, variable $\varepsilon _{r}$,
			{\bf (d)} PM4, $a_{\text{Mg}^{2+}}$ = 3.15 $\cdot$ 10$^{-10}$~m, $a_{\text{Cl}^{-}}$ = 1.81 $\cdot$ 10$^{-10}$~m, variable $\varepsilon _{r}$.}
		\label{fig2}
	\end{center}
\end{figure}
\begin{figure}[!t]
	\begin{center}
		\centerline{\includegraphics[width=0.35\textwidth]{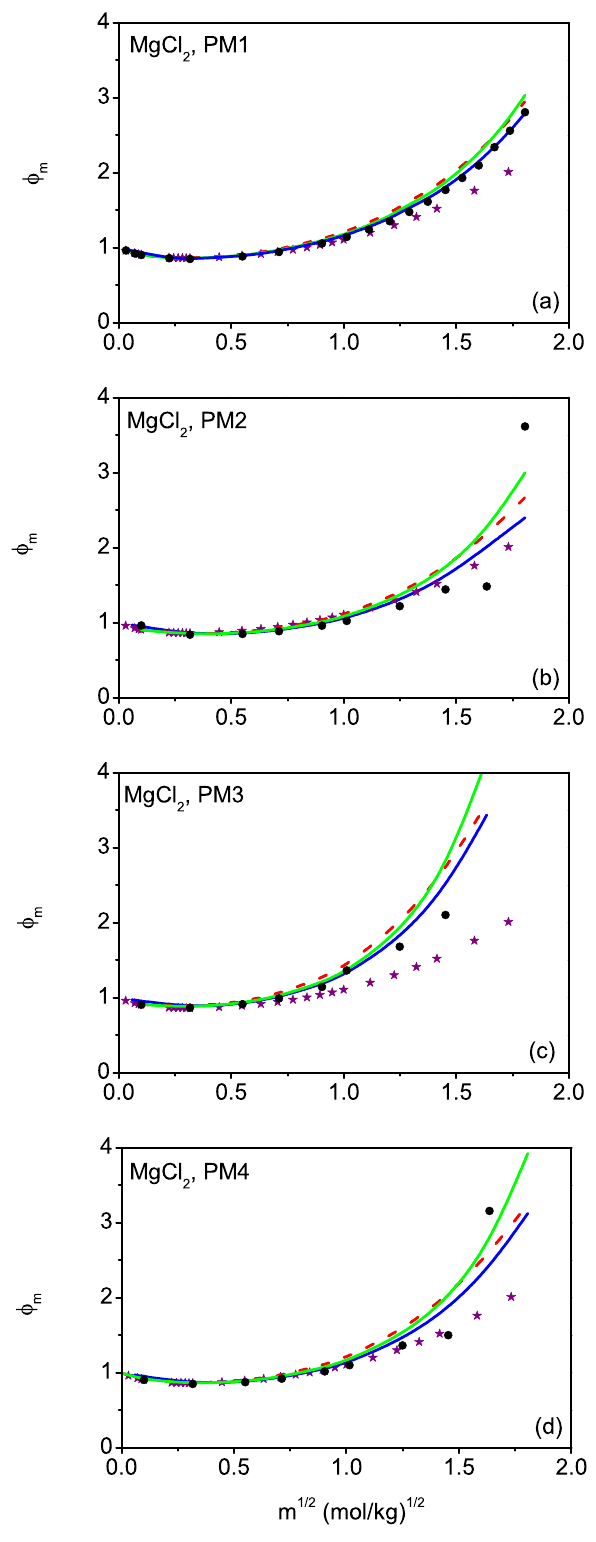}}
		\caption{(Colour online) MPB, SPB, MSA, and MC osmotic coefficient $\phi _{m}$
			(in molal scale) as function of square root of concentration $m^{1/2}$ (in (mol/kg)$^{1/2}$)
			for MgCl$_{2}$ using the PM with a concentration dependent dielectric constant $\varepsilon _{r}$.
			In all of the sub-figures, dashed red line SPB, solid green line MPB, solid blue line MSA, black filled circles MC,
			purple stars experiment \cite{Barthel}. Rest of the legend as in figure~\ref{fig2}.}
		\label{fig3}
	\end{center}
\end{figure}

    The physical model outlined above was treated by MC simulations, and by
the SPB, the MPB, and the MSA theories. Experimental results for the osmotic
and activity coefficients are usually expressed in the molal scale rather than
in the  molar scale used in simulations and theories. We  convert our
(molar scale) data to molal scale using standard conversion relations.
To be consistent with the conventional nomenclature in the literature,
we  use the symbols $c$ and $m$ to denote solution concentration in
mol/dm$^{3}$ (molar scale) and mol/kg (molal scale), respectively, and the
symbols $y$ and $\gamma $ to denote the activity coefficient in molar scale
and in molal scale, respectively. To distinguish the osmotic coefficient
in the two scales, we simply use $\phi _{c}$ and $\phi _{m}$.

    Conversion of ``molarity'' to ``molality'' and vice versa can be achieved
using the relation \cite{Robinson}
\begin{equation}
m=\frac{c}{d-10^{-3}cM},
\end{equation}
where $M$ is the molar mass of the solute and $d$ is the density
of the solution.

    Similarly, conversion of mean activity in molar scale to mean activity
in molal scale and vice versa is accomplished by the relation \cite{Robinson}
\begin{equation}
\gamma _{\pm}= \frac{c}{m}\frac{y_{\pm}}{d_{0}},
\end{equation}
with $d_{0}$ being the density of the pure solvent.
An analogous equation exists for the conversion of $\phi _{c}$ to~$\phi _{m}$.

\subsubsection{MC simulations}

    The MC simulations were performed in a canonical (NVT) ensemble using a
cubic box subject to periodic boundary conditions. The target concentration
of the electrolyte solution was attained by trial-and-error varying the length
of the edge of the box, while keeping the number of particles fixed. The osmotic
coefficients were obtained from the calculated pair correlation (radial
distribution) functions and the virial expression \cite{Outh}, while the activity
coefficients were evaluated through a modified Widom particle insertion technique
\cite{Svensson,Abbas}. { Specifically, the Widom method \cite{Abbas} states that
a non-perturbing particle of species \emph{s} inserted at a random position \emph{r}
will have the individual activity coefficient $y_{s}$ given by
\begin{equation}
\ln y_{s} = -\ln \langle\left\{\exp\left[-\beta \Delta U_{s}(r)\right]\right\}\rangle.
\end{equation}
 Here, $\beta = 1/(k_{\text{B}}T)$, with $k_{\text{B}}$ being the Boltzmann's constant and $T$ being the
absolute temperature. The exponential term enclosed in brackets is the ensemble average of
the energy change, $\Delta U_{s}$ due to the particle addition. This method
provides a direct calculation of the chemical potential. However, the original Widom
method becomes less accurate when dealing with ionic systems of finite sizes,
since the addition of a charged particle will violate electroneutrality in the MC cell.
This effect can be considerably reduced by using a charge rescaling method.
Charge rescaling is a simple method to re-establish electroneutrality in the
computation cell. This method has shown good results for symmetric as well as for
asymmetric electrolytes. The mean activity coefficient $y_{\pm}$ is now obtained
from the $y_{s}$, following the standard practice \cite{McQuarrie}.

    The contact value of the pair correlation function $g_{st}(a_{st})$
between two ions  $s$ and $t$ needed
in the calculation of the $\phi _{c}$ is obtained by a second order polynomial fit to
$g_{st}(r)$ for $r$ close to contact. Further details of the MC method can be found in
reference \cite{Abbas} and in some relevant literature cited there.

Explicitly, the expression used for $\phi _{c}$ is taken from reference \cite{Rasaiah}
[see equation (4.4)], which can be written, using our notations (see also
references \cite{Outh3,Outh4,Outh4ol}), as
\begin{equation}
\phi _{c}=1+\frac{2\piup}{3\rho}\sum_{s}\sum_{t}\rho _{s}\rho _{t}g_{st}(a_{st})a_{st}^{3} + \frac{\beta E}{3\rho },
\label{eq2.5}
\end{equation}
where $E$ is the (MC) excess energy, $\rho_{s}$ is the mean number
density of ions of type $s$, and $\rho = \sum_{s}\rho _{s}$ with the summation being over
the number of ionic species. In the simulation calculations for $\phi_{c}$, the
excess energy term $E$ was evaluated via its representation in terms of the $g_{st}(r)$
rather than the mean electrostatic potential \cite{Outh}.}

    As a check on the numerics we repeated the calculations in the grand
canonical ensemble. The results for the thermodynamic data from the two ensembles
were within statistical error of each other. We used a large number of particles
of the order $\sim $10$^{3}$ and generated a large number of MC configurations
of the order $\sim $10$^{8}$. The first 10$\%$ were used for system equilibration,
while the rest were used in taking statistics. Concentration dependence of the
$\varepsilon _{r}$ was ensured by using experimentally measured values of the
quantity at different concentrations for the salt solutions treated.

\subsubsection{SPB and MPB theories}

    The SPB and MPB constitute potential formulations of the theory to describe
electrolyte solutions. An important difference between the classical, non-linear
Poisson-Boltzmann (PB) theory and the SPB, MPB is that while in the former the
pair correlation function $g_{ij}(r)$ between two ions $i$ and $j$ is asymmetric,
for instance, $g_{ij}(r)\neq g_{ji}(r)$, with respect to an interchange of indices,
for any asymmetry in the system, the latter two are devoid of this defect. The
SPB is still a mean-field theory, although, as we will see later, some hard core
effects are taken into account through the exclusion volume envelope. The MPB
accounts for both the ionic exclusion volume effects and for the fluctuation potential
missing in the SPB. The development of SPB and MPB formalisms over the years have
been detailed in the literature (see for example~\cite{Outh3,Outh4,Outh4ol,Ulloa}).
We will therefore be brief and simply outline the principal equations involved
in the theories.

    The starting point is the Poisson's equation
\begin{equation}
\nabla ^{2}\psi _{s}(1;2)=-\frac{|e|}{\varepsilon _{0}\varepsilon_{r}}\sum _{t}Z_{t}\rho _{t}g_{st}(r_{st}).
\label{eq2.6}
\end{equation}
 Here, $\psi _{s}(1;2)$ is the mean electrostatic potential about an ion $s$ at ${\bf r}_{1}$
at a field point ${\bf r}_{2}$, $g_{st}$ is the pair correlation function for the ion pair
$s$ and $t$ with separation $r_{st}=|{\bf r}_{1}-{\bf r}_{2}|$. Note that
we will be dealing here with single electrolytes so that $s = i, j$ ($i, j$ being $+,-$ or $-, +$).

   For the PM, in the SPB theory, the symmetrized $g_{st}(r)$ ($r = r_{st}$) is developed as
\begin{equation}
g_{st}(r)=g_{st}^{0}(r)\exp\left\{-\frac{\beta e}{2}\left[Z_{s}(\psi _{t}(r)+\psi _{t}^{0}(r))+Z_{t}(\psi _{s}(r)+
\psi _{s}^{0}(r))\right]\right\},
\label{eq2.7}
\end{equation}
where $\psi _{s}^{0}(r) = \psi _{s}(r;Z_{s} = 0)$ is the (discharged) potential at
a distance $r$ from the discharged ion $s$. The $g_{st}^{0}$ in the above equation is the
exclusion volume term, which is the pair correlation between two discharged ions in a sea of
fully charged ions.

    In the MPB, the $g_{st}(r)$ is given by
\begin{equation}
g_{st}(r)=g_{st}^{0}(r)\exp\left\{-\frac{\beta e}{2}\left[Z_{s}(L_{t}(u_{t})+L_{t}(u_{t}|Z_{t}=0))+Z_{t}(L_{s}(u_{s})+
L_{s}(u_{s}|Z_{s}=0))\right]\right\}.
\label{eq2.8}
\end{equation}
Here, $u_{s}=r\psi _{s}(r)$, and assuming without loss of generality
that $a_{i} \leqslant a_{j}$, we have
\begin{equation}
L_{s}(u)= \frac{1}{2r(1+\kappa a_{is})}\left[u(r+a_{is})+u(r-a_{is})+ \kappa \int_{r-a_{is}}^{r+a_{is}}u(R)\rd R\right],
\label{eq2.9}
\end{equation}
 with the ion $i$ having the smallest radius.
The Debye-H\"{u}ckel parameter $\kappa $ is given by
\begin{equation}
\kappa = \left[\frac{e^{2}\beta}{\varepsilon_{0}\varepsilon_{r}}\sum_{s} Z_{s}^{2} \rho_{s}\right]^{1/2}.
\label{eq2.10}
\end{equation}

    The exclusion volume term  $g_{is}^{0}$ was approximated by the Percus-Yevick
uncharged pair distributions~\cite{Hansen,leonardhendersonbarker,grundkehenderson} and
their corrections due to Verlet and Weis \cite{verletweis}.
{ The equations (\ref{eq2.6}) and (\ref{eq2.7}) constitute the SPB equation, while the equations
(\ref{eq2.6}),(\ref{eq2.8})--(\ref{eq2.10}) constitute the MPB equation.

    To calculate the $\phi _{c}$ and $y_{\pm}$ in the SPB and MPB theories,
we have used the virial route and the Guntelberg charging route, respectively
\cite{Outh4,Outh4ol,Molero}. These routes represent the optimum thermodynamic ones for
these theories.

    The expression for $\phi _{c}$ used in both the SPB and MPB is that given
above, equation~(\ref{eq2.5}),  where the excess energy $E$ in terms of the mean electrostatic potential is
\begin{equation}
E=\frac{e}{2}\sum_{s}Z_{s}\rho _{s}\left[\psi _{s}(a_{is})-\frac{eZ_{s}}{4\piup \varepsilon_{0}\varepsilon_{r}a_{is}}\right].
\label{eq2.11}
\end{equation}
The individual ionic activity is written in terms of the hard sphere (HS) and electrical ({el}) parts.
\begin{equation}
\ln y_{s}=\ln y_{s}^{\text{HS}}+\ln y_{s}^{\text{el}},
\end{equation}
\begin{equation}
\ln y_{s}^{\text{el}}=e\beta Z_{s}\int _{0}^{1}\rd\lambda \left[\psi _{s}(a_{is},\lambda)-\frac{\lambda eZ_{s}}{4\piup \varepsilon_{0}\varepsilon_{r}a_{is}}\right],
\label{eq2.13}
\end{equation}
with the hard sphere contribution being obtained from the Ebeling and Scherwinski results
\cite{Ebeling}. As before with MC, the mean activity $y_{\pm}$ is obtained from the $y_{s}$
in the usual manner \cite{McQuarrie}.} We note that since $a_{i}$ is fixed as the smallest ion [see text below equation~(\ref{eq2.9})], so in the right hand side of equation~(\ref{eq2.11})
whenever we have $a_{is}$, $s$ is regarded as variable and $ a_{is}$ could be either $a_{ii}$ or $a_{ij}$.
$E$ is simply the excess energy independent of $s$, while in equation (\ref{eq2.13}) $\ln y_{s}^{\text{el}}$ is the individual activity (electrical part)
for ions of type $s$. With equal ion sizes the expressions for $E$ and $\ln y_{s}^{\text{el}}$ reduce to those in reference \cite{Outhwaite48}.
\subsubsection{MSA}
\begin{figure}[!t]
	\begin{center}
		\centerline{\includegraphics[width=0.35\textwidth]{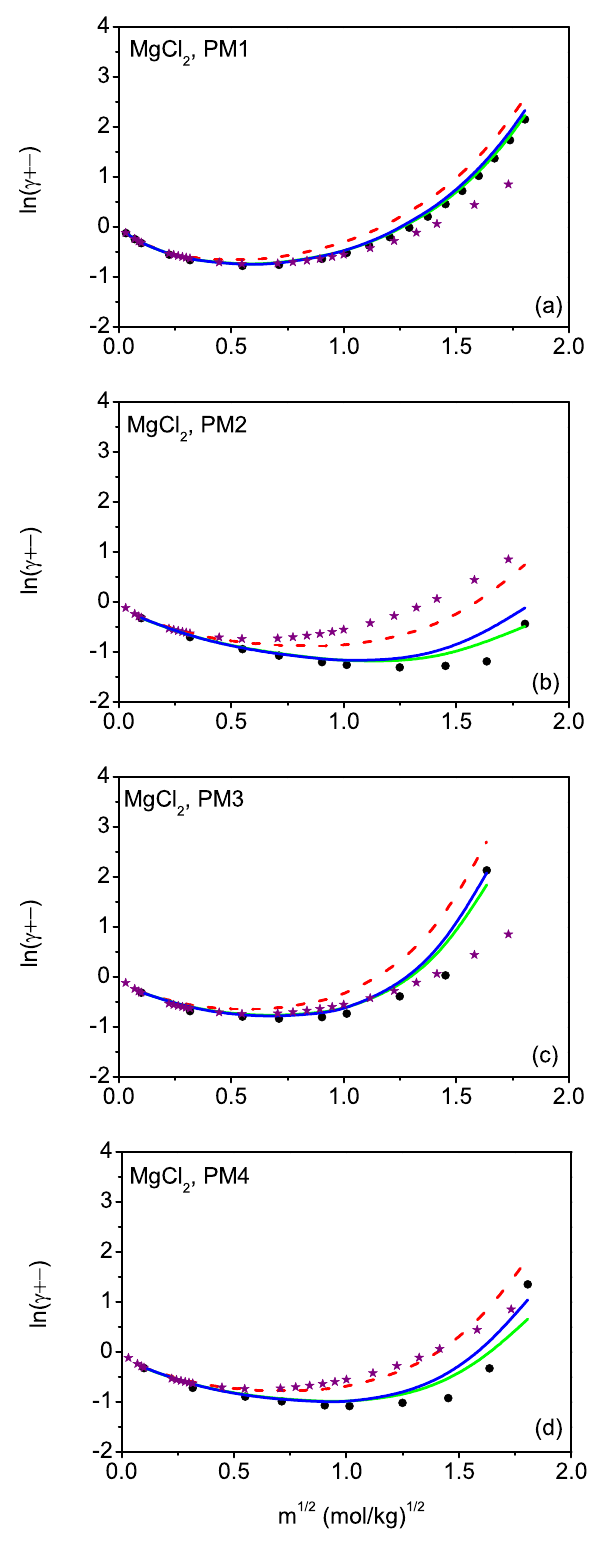}}
		\caption{(Colour online) MPB, SPB, MSA, and MC mean activity coefficient $y_{\pm}$
			(in molar scale) as function of square root of concentration $c^{1/2}$ (in (mol/dm$^{3}$)$^{1/2}$)
			for MgCl$_{2}$ using the PM with a concentration dependent dielectric constant $\varepsilon _{r}$.
			In all of the sub-figures, dashed red line SPB, solid green line MPB, solid blue line MSA, black
			filled circles MC. Rest of the legend as in figure~\ref{fig2}.}
		\label{fig4}
	\end{center}
\end{figure}
\begin{figure}[!t]
	\begin{center}
		\centerline{\includegraphics[width=0.35\textwidth]{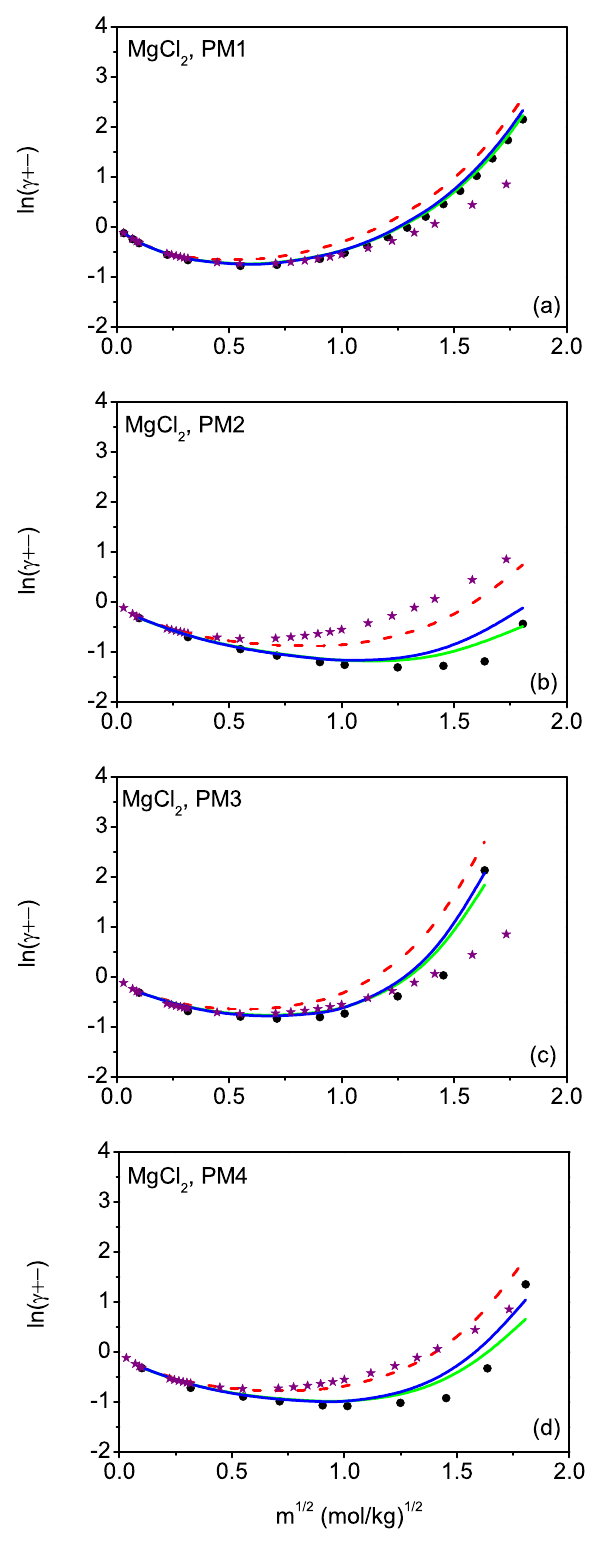}}
		\caption{(Colour online) MPB, SPB, MSA, and MC mean activity coefficient $\gamma _{\pm}$
			(in molal scale) as function of square root of concentration $m^{1/2}$ (in (mol/kg)$^{1/2}$)
			for MgCl$_{2}$ using the PM with a concentration dependent dielectric constant $\varepsilon _{r}$.
			In all of the sub-figures, dashed red line SPB, solid green line MPB, solid blue line MSA, black filled circles MC,
			purple stars experiment \cite{Barthel}. Rest of the legend as in figure~\ref{fig2}.}
		\label{fig5}
	\end{center}
\end{figure}

The MSA is a linear, integral equation theory with the advantage that it is
analytically tractable~\cite{Blum, Blum2}. The thermodynamic quantities of interest
such as the internal energy, the osmotic coefficient, and the activity coefficients
have got closed analytic forms \cite{Blum3, Sanchez}.

    The starting point here is the homogeneous Ornstein-Zernike (OZ) equation
\begin{equation}
h_{st}(r_{st})=c_{st}(r_{st})+\sum_{m}\int h_{sm}(r_{sm})c_{mt}(r_{mt})\rd{\bf r}_{3},
\end{equation}
where $r_{sm}=|{\bf r}_{1}-{\bf r}_{3}|$, and $h_{st}(=g_{st}-1)$ and $c_{st}$ are
the total and direct correlation functions, respectively.

    The OZ equation together with the following (MSA) closure relations comprise the MSA theory,
\begin{equation}
g_{ij}(r)=0 ,  \quad           r < a_{ij},
\label{eq2.15}
\end{equation}
\begin{equation}
c_{ij}(r)=-\beta u_{ij}(r), \quad   r > a_{ij}.
\label{eq2.16}
\end{equation}
Note that equation (\ref{eq2.15}) is an exact condition, while
equation (\ref{eq2.16}) is an approximation. For the MSA, the energy route
is the most accurate thermodynamic route and hence this was employed
\cite{Blum3,Sanchez} to evaluate the $\phi _{c}$ and $y_{\pm}$.

\section{Results}
\begin{figure}[!t]
	\begin{center}
		\centerline{\includegraphics[width=0.35\textwidth]{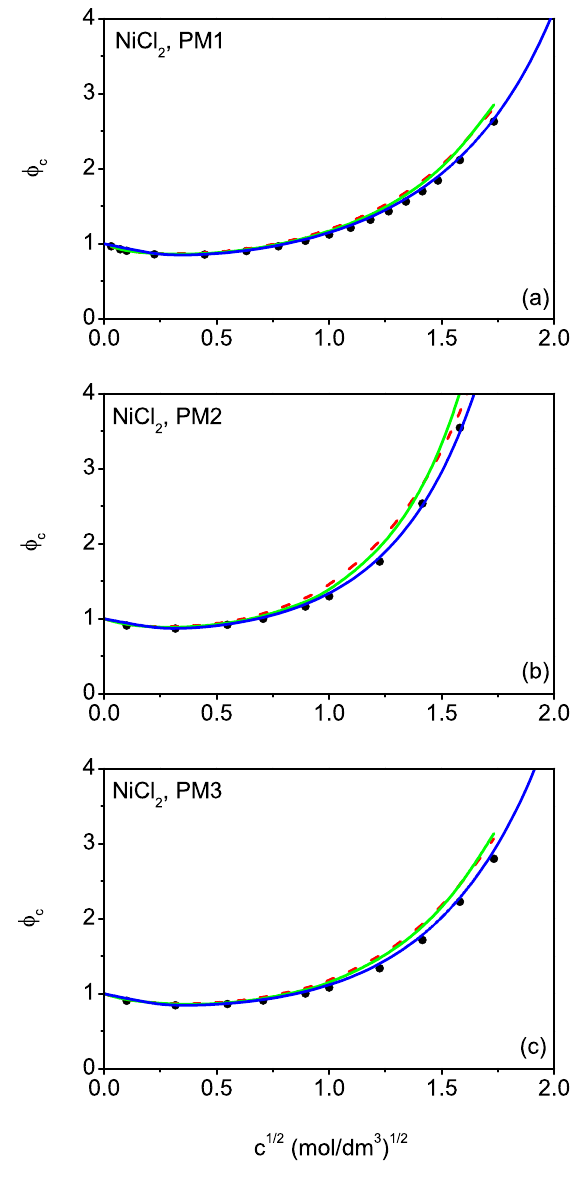}}
		\caption{(Colour online) MPB, SPB, MSA, and MC osmotic coefficient $\phi _{c}$
			(in molar scale) as function of square root of concentration $c^{1/2}$ (in (mol/dm$^{3}$)$^{1/2}$)
			for NiCl$_{2}$ using the PM with a concentration dependent dielectric constant $\varepsilon _{r}$.
			In all of the sub-figures, dashed red line SPB, solid green line MPB, solid blue line MSA, black filled circles MC;
			{\bf (a)} PM1, $a_{\text{Ni}^{2+}}$ = 2.80 $\cdot$ 10$^{-10}$~m, $a_{\text{Cl}^{-}}$ = 1.81 $\cdot$ 10$^{-10}$~m, fixed $\varepsilon _{r}$,
			{\bf (b)} PM2, $a_{\text{Ni}^{2+}}$ = 3.40 $\cdot$ 10$^{-10}$~m, $a_{\text{Cl}^{-}}$ = 1.81 $\cdot$ 10$^{-10}$~m, variable $\varepsilon _{r}$,
			{\bf (c)} PM3, $a_{\text{Ni}^{2+}}$ =~2.90~$\cdot$~10$^{-10}$~m, $a_{\text{Cl}^{-}}$ = 1.81 $\cdot$ 10$^{-10}$~m, variable $\varepsilon _{r}$.}
		\label{fig6}
	\end{center}
\end{figure}
\begin{figure}[!t]
	\begin{center}
		\centerline{\includegraphics[width=0.35\textwidth]{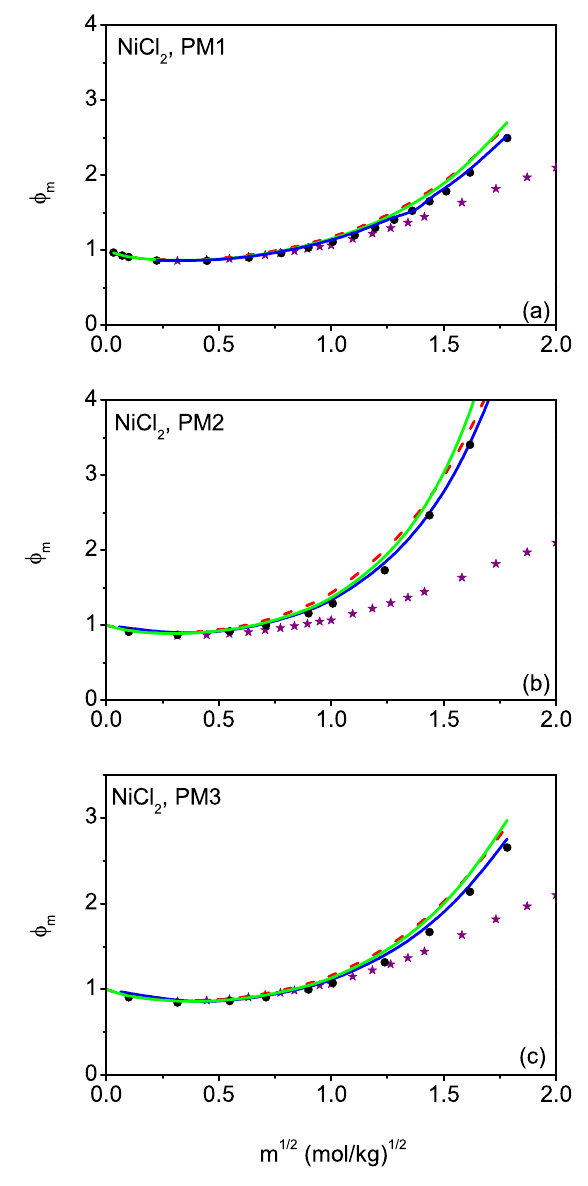}}
		\caption{(Colour online) MPB, SPB, MSA, and MC mean activity coefficient $\gamma _{\pm}$
			(in molal scale) as function of square root of concentration $m^{1/2}$ [in (mol/kg)$^{1/2}$]
			for NiCl$_{2}$ using the PM with a concentration dependent dielectric constant $\varepsilon _{r}$.
			In all of the sub-figures, dashed red line SPB, solid green line MPB, solid blue line MSA, black filled circles MC,
			purple stars experiment \cite{Barthel}. Rest of the legend as in figure~\ref{fig6}.}
		\label{fig7}
	\end{center}
\end{figure}

    The SPB and MPB equations were solved numerically using a previously used
quasi-linearization iteration procedure \cite{Bellman}. This is a robust technique
which has proved useful in earlier works (see for example~\cite{Martinez,Molero,
Outh3,Outh4,Outh4ol}). The MSA results for the osmotic and activity coefficients were obtained
from their analytic expressions \cite{Blum,Blum3,Sanchez}. The necessary hard sphere
contributions to the activity coefficients were obtained from the work of Ebeling
and Scherwinski \cite{Ebeling}.

    We present the results here for $\phi _{c}(\phi _{m})$ and $y_{\pm}(\gamma _{\pm})$
for MgCl$_{2}$ and NiCl$_{2}$ solutions with concentration dependent $\varepsilon _{r}$
(cf. equations (\ref{eq1.1}) and (\ref{eq1.2})) at $T = 298$ K. Some results at fixed $\varepsilon _{r} = 78.36$
have also been included for comparison purposes. For each of the salts, different
combinations of cation and anion radii were used. In the MC simulations,
the adopted strategy of keeping the anion size equal to its crystallographic value,
while the cation size is optimized to fit the experimental data, was based on the
dielectric relaxation spectroscopic work on electrolyte solutions by Buchner and
Hefter \cite{Buchner}. They have shown, for example, that Cl$^{-}$ ions have
vanishingly small hydration shells around them, whereas, the Mg$^{2+}$ ions
display strong hydration. We have used four such sets for MgCl$_{2}$:

\indent (i)\hspace{0.14cm}   PM1 $a_{\text{Mg}^{2+}}$ = 2.95 $\cdot$ 10$^{-10}$ m, $a_{\text{Cl}^{-}}$ = 1.81 $\cdot$ 10$^{-10}$ m, fixed $\varepsilon _{r}$,

\indent (ii)\hspace{0.075cm}  PM2 $a_{\text{Mg}^{2+}}$ = 2.95 $\cdot$ 10$^{-10}$ m, $a_{\text{Cl}^{-}}$ = 1.81 $\cdot$ 10$^{-10}$ m, variable $\varepsilon _{r}$,

\indent (iii) PM3 $a_{\text{Mg}^{2+}}$ = 3.50 $\cdot$ 10$^{-10}$ m, $a_{\text{Cl}^{-}}$ = 1.81 $\cdot$ 10$^{-10}$ m, variable $\varepsilon _{r}$,

\indent (iv) PM4 $a_{\text{Mg}^{2+}}$ = 3.15 $\cdot$ 10$^{-10}$ m, $a_{\text{Cl}^{-}}$ = 1.81 $\cdot$ 10$^{-10}$ m, variable $\varepsilon _{r}$

\indent and three sets for NiCl$_{2}$:

\indent (i) \hspace{0.12cm} PM1 $a_{\text{Ni}^{2+}}$ = 2.80 $\cdot$ 10$^{-10}$ m, $a_{\text{Cl}^{-}}$ = 1.81 $\cdot$ 10$^{-10}$ m, fixed $\varepsilon _{r}$,

\indent (ii) \hspace{0.04cm} PM2 $a_{\text{Ni}^{2+}}$ = 3.40 $\cdot$ 10$^{-10}$ m, $a_{\text{Cl}^{-}}$ = 1.81 $\cdot$ 10$^{-10}$ m, variable $\varepsilon _{r}$,

\indent (iii) \hspace{0.0012cm} PM3 $a_{\text{Ni}^{2+}}$ = 2.90 $\cdot$ 10$^{-10}$ m, $a_{\text{Cl}^{-}}$ = 1.81 $\cdot$ 10$^{-10}$ m, variable $\varepsilon _{r}$.

 \noindent These radii were also used in the SPB, MPB, and MSA applications.
The best-fit to the experimental data~\cite{Lobo,Stokes,Stokes2,Stokes3} is achieved by
PM4 for MgCl$_{2}$ and by PM3 for NiCl$_{2}$.

    Figures~\ref{fig2} and \ref{fig3} show the osmotic coefficient of MgCl$_{2}$ in molar scale
(as a function of $c^{1/2}$) and in molal scale (as a function of $m^{1/2}$),
respectively, for the four PM's. The theoretical curves are all consistent with
each other across the four models. In general, the theories predict the simulation
data well up to $c\sim $ 1.6 mol/dm$^{3}$. The SPB curves tend to deviate more, as
expected, at higher concentrations when the neglect of the fluctuation potential
becomes more consequential. For the PM2--PM4, the MC data show a little noise
beyond $c\sim $ 2.25 mol/dm$^{3}$. Indeed, in PM4 at $c =$ 3 mol/dm$^{3}$, the
MC $\phi _{c} =$ 26.52, which may well be an outlier. The corresponding curves
in molal scale in figure~\ref{fig3} are compared with the experimental MgCl$_{2}$ $\phi _{m}$
data \cite{Barthel}. The simulations and the theoretical predictions appear
relatively closer to the experimental data in PM1 and in PM2 than they do in PM3 and PM4.
In the latter two cases the simulations begin to show deviations from the experiments
beyond $m \sim$ 1 mol/kg.

    We now turn to figures~\ref{fig4} and \ref{fig5}, which illustrate the mean activity coefficient
of MgCl$_{2}$, again in molar and molal scales, respectively. Apropos of what we saw
in figure~\ref{fig2}, the theories, with the exception of the SPB, reproduce the MC data
closely for PM1. For PM2--PM4, notwithstanding some statistical noise in the simulation data at
higher concentrations, the theoretical curves begin to deviate
beyond $c\sim $~1.6~mol/dm$^{3}$. The deviation of SPB is more than that of MSA or MPB.
The trends of these curves in figure~\ref{fig5} are similar, with the simulations and the theories
being relatively closer to the experimental data in PM1 and PM3 up to $\sim$ 1.6 mol/kg.
It is of interest to note that the simulations and theories overestimate the experiments
in figures~\ref{fig1} and \ref{fig3}, but underestimate them in PM2 and PM4.
\begin{figure}[!t]
	\begin{center}
		\centerline{\includegraphics[width=0.4\textwidth]{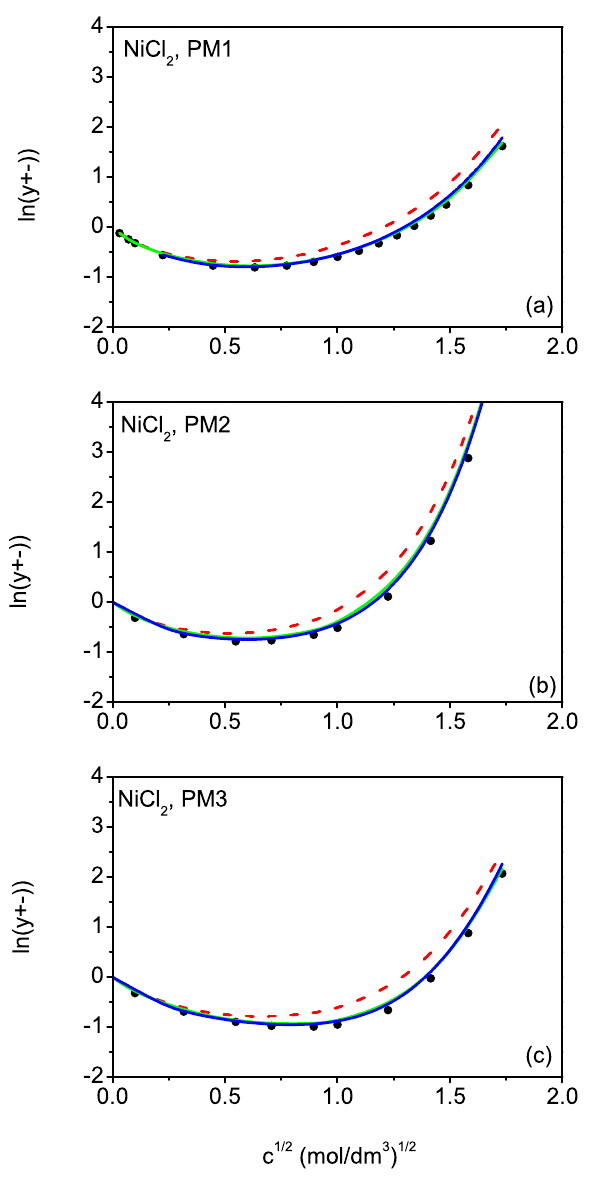}}
		\caption{(Colour online) MPB, SPB, MSA, and MC mean activity coefficient $y_{\pm}$
			(in molar scale) as function of square root of concentration $c^{1/2}$ (in (mol/dm$^{3}$)$^{1/2}$)
			for NiCl$_{2}$ using the PM with a concentration dependent dielectric constant $\varepsilon _{r}$.
			In all of the sub-figures, dashed red line SPB, solid green line MPB, solid blue line MSA, black filled circles MC.
			Rest of the legend as in figure~\ref{fig6}.}
		\label{fig8}
	\end{center}
\end{figure}
\begin{figure}[!t]
	\begin{center}
		\centerline{\includegraphics[width=0.4\textwidth]{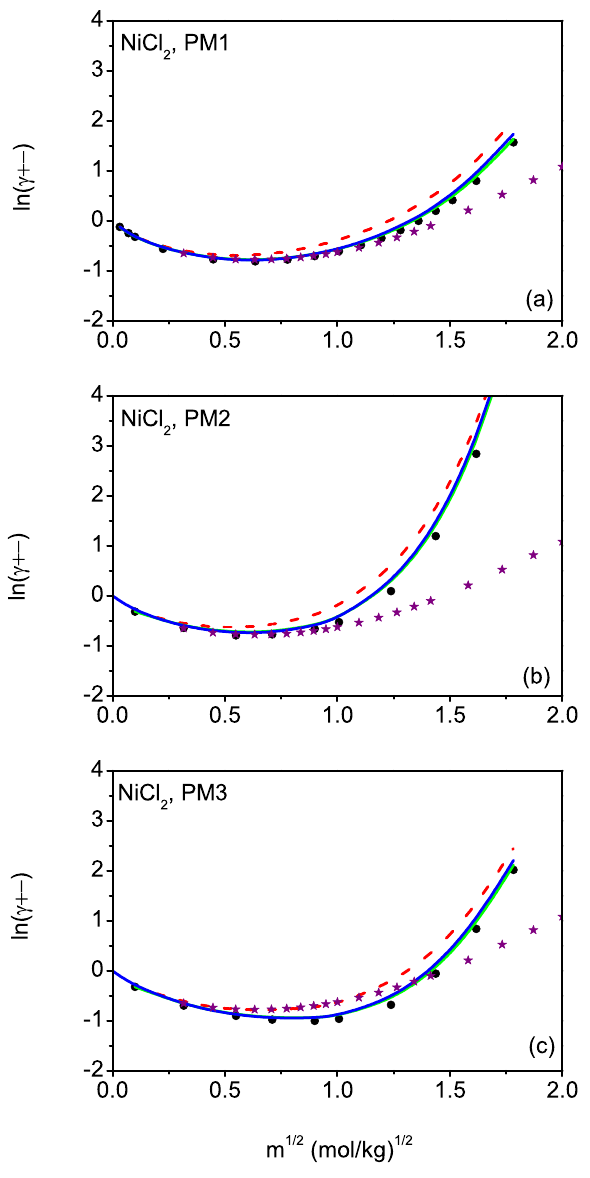}}
		\caption{(Colour online) MPB, SPB, MSA, and MC mean activity coefficient $\gamma _{\pm}$
			(in molal scale) as function of square root of concentration $m^{1/2}$ (in (mol/kg)$^{1/2}$)
			for MgCl$_{2}$ using the PM with a concentration dependent dielectric constant $\varepsilon _{r}$.
			In all of the sub-figures, dashed red line SPB, solid green line MPB, solid blue line MSA, black filled circles MC,
			purple stars experiment \cite{Barthel}. Rest of the legend as in figure~\ref{fig6}.}
		\label{fig9}
	\end{center}
\end{figure}

    Turning now to the results for NiCl$_{2}$, we show the osmotic coefficient in figure~\ref{fig6}
(molar scale) and in figure~\ref{fig7} (molal scale). In figure~\ref{fig7} the MSA and MPB predictions match
the simulations rather well for all of the three PM's, with the MSA in particular being
almost quantitative with the MC data. It is noted that at higher concentrations, the agreement
of these curves with the simulations is better than that seen before with MgCl$_{2}$. The
SPB curve begins to deviate from the MC data from about $c\sim $ 0.25 mol/dm$^{3}$. The
behaviour of the theoretical curves and the simulation data in the molal scale in figure~\ref{fig7}
is analogous. However, compared to the experimental results, the theoretical curves and the
MC data deviate from about $m\sim $ 1.6 mol/kg in PM1, $m\sim$ 0.6 mol/kg in PM2, and $m\sim$
1 mol/kg in PM3, respectively.

    The pattern of correspondence between the theories and the simulations continues
in figures~\ref{fig8} and \ref{fig9} where the mean activity of NiCl$_{2}$ is presented in molar and molal
scales, respectively. In these figures, the MSA and MPB results are nearly quantitative
with the MC data. The deviation shown by the SPB curve is similar to that seen for the
SPB osmotic coefficient. The reason for the good agreement between the MSA, MPB and the
MC, especially at higher concentrations, can be traced to the behaviour of $\varepsilon _{r}$
versus $c$ for NiCl$_{2}$ seen in figure\ref{fig1}. At concentrations beyond $c\sim$ 1.5 mol/dm$^{3}$,
the $\varepsilon _{r}$ increases leading to an overall decrease in the strength of the electrostatic
interactions, which in turn lead to the observed effects. With regard to the comparative
behaviour of the simulations/theories with the experimental data, the trends are analogous
to that seen with the osmotic coefficient in figure~\ref{fig7}. For PM2, there is a large deviation,
but for PM1 and PM3, the deviation is relatively smaller in magnitude and also sets in at a
higher concentration ($m\sim$ 2 mol/kg).

\section{Conclusions}

    The work is an extension of our earlier work \cite{Quinones2} on the thermodynamics
of symmetric 1:1 valency electrolytes with variable $\varepsilon _{r}$
to asymmetric 2:1 valency MgCl$_{2}$ and NiCl$_{2}$ salts. Higher and
multivalencies pose a sterner theoretical challenge. As in \cite{Quinones2},
we have used experimentally determined empirical formula for $\varepsilon _{r}$
as a function of concentration for both of these salts.

    We have utilized the statistical mechanical formalisms of SPB, MPB,
and MSA in conjunction with MC simulations, which are usually accepted
as \emph{gold standard} in this field, in order to compute the osmotic and activity
coefficients of these salts. We have further compared the simulation data
and the theoretical predictions with experimental results for these quantities
from the literature \cite{Barthel}.

    Overall, the theoretically predicted results for $\phi _{c}$ and $y_{\pm}$
with the exception perhaps of that of the SPB, compare well with the corresponding
simulation data. In some cases, the MSA and the MPB results are semi-quantitative or
better. However, there are discrepancies at higher concentrations with the onset
of such deviations occurring at lower concentrations for the SPB. For some ionic
radii, combinations of MgCl$_{2}$, the MC data show some dispersions (beyond normal
statistical error) at higher concentrations. Admittedly, in \cite{Quinones2}
the theoretical results were seen to be broadly closer to the simulations than
they are in the present calculations. This is clearly due to the asymmetric 2:1
valency salts being treated here as opposed to the symmetric 1:1 salts in \cite{Quinones2}.
With regard to the comparison with experimental data, the theories and
simulations follow them reasonably well at low to intermediate concentrations.
Although there are deviations at higher densities of the salts, we should
emphasize that the agreement is good up to a respectable $m\sim $1 mol/kg.

    As indicated earlier, at higher concentrations, the MSA or the MPB thermodynamics
reveal a relatively better agreement with the MC data for NiCl$_{2}$ rather than for
MgCl$_{2}$. This occurs because the MgCl$_{2}$ $\varepsilon _{r}$ displays a
continual dielectric decrement with concentration, whereas the NiCl$_{2}$
$\varepsilon _{r}$ actually increases at higher concentrations.
For the latter, this results in dilution of the interionic interactions
as the plasma coupling constant { $\Gamma = e^{2}/(4\piup \varepsilon _{0}\varepsilon _{r}k_{\text{B}}Ta_{+-})$
($a_{-}=a_{\text{Cl}^{-}}, a_{+}=a_{\text{Mg}^{2+}}$ or $a_{\text{Ni}^{2+}}$)} decreases,
which tends to negate somewhat the effects of higher valency.

{ The comparisons of simulation and theory with the experimental
data show that the empirical relations~(\ref{eq1.1}) and (\ref{eq1.2}) fail at the higher
molal concentrations. A possible improvement to these empirical relations is
indicated by the work of Barthel et al. \cite{Barthel2} for non-aqueous electrolyte
solutions. These authors extended their relation, equation (70), analogous to ours
by using a Pade-approximant to correct for saturation effects at high concentrations.}

{ In view of the steep increases for some of the osmotic and
mean activity coefficients that occur at higher concentrations, it is tempting
to wonder whether this could somehow be related to underscreening or charge inversion.
At these higher concentrations, underscreening occurs, with the mean electrostatic
potential displaying a damped oscillatory behavior \cite{Outh5}. The transformation in
the mean potential is associated with the Kirkwood transition (KT) which depends
on both ion size and valence. See for example \cite{Dinpajooh} and the references
therein. The MPB theory is based on the Kirkwood charging process and the KT
identified through some critical value  $\alpha _{c}$ of $\alpha = \kappa a_{ij}$.
For the RPM, the present MPB does not distinguish between symmetric and unsymmetric
valencies, only varying the ion size \cite{Thomlinson}. Considering solely the best
fit values of PM4 for MgCl$_{2}$ and PM3 for NiCl$_{2}$, the critical values of
$\alpha _{c}$ are $\sim $0.91 and $\sim $0.98, respectively. These values of
$\alpha _{c}$ correspond to $c \sim$ 0.10 mol/dm$^{3}$ and $c \sim$ 0.13 mol/dm$^{3}$
for the two salts, respectively. At the highest concentration of 3 mol/dm$^{3}$,
the respective values of $\alpha$ for MgCl$_{2}$ and NiCl$_{2}$ are 7.44 and 4.55.
These values show that the rapid increase in the thermodynamic properties at the higher
concentrations occurs when there is underscreening.}
	
    From a theoretical perspective, an analysis that takes into account the discreteness
of the solvent rather than treating it as a continuum would be more appealing. The use of hard
spheres as solvent saw a limited success since a dielectric constant needed to be assigned
separately \cite{Bhuiyan}. Similarly, a dipolar solvent proved less than appropriate since the
predicted $\varepsilon _{r}$ proved to be low for aqueous electrolytes~\cite{Kournopoulos}. Under
the circumstances, the recourse to experimentally measured $\varepsilon _{r}$ represents a more
feasible way forward. Obvious advantages are the facts that the $\varepsilon _{r}$ being
used are more realistic and the process is more transparent.

\section*{Acknowledgements}

\hspace{0.14 cm} AOQ acknowledges a graduate studentship from the Decanato de Estudios Graduados
e Investigaci\'{o}n (DEGI), University of Puerto Rico-Rio Piedras Campus.


\ukrainianpart

\title
{┬яышт чь│ээю┐ т│фэюёэю┐ ф│хыхъЄЁшўэю┐ яЁюэшъэюёЄ│ эр ЄхЁьюфшэрь│ъє рёшьхЄЁшўэю┐ трыхэЄэюёЄ│ тюфэшї ёюыхщ
}
\author[A. O. ╩тiэюэхё, ╟. └ссрё, ╩. ┬. ╬єётхщЄ, ╦. ┴. ┴єi э]{
	A. O. ╩тiэюэхё\refaddr{label1}, ╟. └ссрё\refaddr{label2}, ╩. ┬. ╬єётхщЄ\refaddr{label3},
	 ╦. ┴. ┴єi э\refaddr{label1}  }

\addresses{
	\addr{label1} ╦рсюЁрЄюЁi  ЄхюЁхЄшўэю┐ Їiчшъш, ╘iчшўэшщ Їръєы№ЄхЄ єэiтхЁёшЄхЄє ╧єхЁЄю-╨iъю, 17 └тхэiфр ╙эiтхЁёшфрф, STE 1701, ╤рэ ╒єрэ, ╧єхЁЄю-╨iъю 00925-2537, ╤╪└
	\addr{label2} ╘ръєы№ЄхЄ ї│ь│┐ Єр ьюыхъєы Ёэю┐ с│юыюу│┐ єэ│тхЁёшЄхЄє ехЄхсюЁур, ╩хэ│эурЁфхэ 4, SE-41296, ехЄхсюЁу, ╪тхЎ│ 
	\addr{label3} ┬ш∙р °ъюыр ьрЄхьрЄшъш Єр Ї│чшъш, єэ│тхЁёшЄхЄ ╪хЇЇ│ыфр, ╪хЇЇ│ыф S3 7RH, ┬хышъюсЁшЄрэ│ 
}

\makeukrtitle

\begin{abstract}

	┼ъёяхЁшьхэЄры№эю тшчэрўхэ│ хья│Ёшўэ│ ЇюЁьєыш фы 
	чрыхцэю┐ т│ф ъюэЎхэЄЁрЎ│┐ т│фэюёэю┐ ф│хыхъЄЁшўэю┐ яЁюэшъэюёЄ│ тюфэшї Ёючўшэ│т MgCl$_{2}$ │ NiCl$_{2}$ тшъюЁшёЄютє■Є№ё  фы  ЁючЁрїєэъє юёьюЄшўэюую ъюхЇ│Ў│║эЄр Єр ёхЁхфэ№юую ъюхЇ│Ў│║эЄр ръЄштэюёЄ│ Ўшї ёюыхщ є °шЁюъюьє ф│рярчюэ│ ъюэЎхэЄЁрЎ│щ. ╤шёЄхьш ьюфхы■■Є№ё  чр фюяюьюую■ яЁшь│Єштэю┐ ьюфхы│ хыхъЄЁюы│Є│т Єр рэры│чє■Є№ё  эр юёэют│ ёшьхЄЁшўэю┐ ЄхюЁ│┐ ╧єрёёюэр-┴юы№Ўьрэр, ьюфшЇ│ъютрэю┐ ЄхюЁ│┐ ╧єрёёюэр-┴юы№Ўьрэр, ёхЁхфэ№ю-ёЇхЁшўэюую эрсышцхээ  Єр ьюфхы■трээ  ╠юэЄх-╩рЁыю. ╟рурыюь, ёхЁхфэ№ю-ёЇхЁшўэх эрсышцхээ  Єр ьюфшЇ│ъютрэр ЄхюЁ│  ╧єрёёюэр-┴юы№Ўьрэр фюсЁх т│фЄтюЁ■■Є№ фрэ│ \emph{хЄрыюээюую} ьюфхы■трээ  фю ъюэЎхэЄЁрЎ│щ $\sim $1,6 ьюы№/фь$^{3}$ │ с│ы№°х, Єюф│  ъ ёшьхЄЁшўэ│ Ёхчєы№ЄрЄш ╧єрёёюэр-┴юы№Ўьрэр ьр■Є№ яхтэ│ т│фїшыхээ , яюўшэр■ўш ч $\sim $0,25 ьюы№/фь$^{3}$. ▀ъ яЁртшыю, │ ьюфхы■трээ , │ ЄхюЁ│┐, яЁштюф Є№ фю яюїшсюъ є яюЁ│тэ ээ│ ч т│фяют│фэшьш хъёяхЁшьхэЄры№эшьш Ёхчєы№ЄрЄрьш яЁш ъюэЎхэЄЁрЎ│ ї яюэрф $\sim $1 ьюы№/ъу.
 	\keywordsяЁшь│Єштэ│ ьюфхы│ хыхъЄЁюы│Є│т, юёьюЄшўэшщ ъюхЇ│Ў│║эЄ Єр ъюхЇ│Ў│║эЄ ръЄштэюёЄ│, ьюфхы■трээ  ╠юэЄх-╩рЁыю, ёшьхЄЁшчютрэ│ Єр ьюфшЇ│ъютрэ│ ЄхюЁ│┐ ╧єрёёюэр-┴юы№Ўьрэр
	
\end{abstract}

\end{document}